\documentclass[12pt]{iopart}
\usepackage{iopams}
\newcommand{\be}{\begin{equation}}
\newcommand{\ee}{\end{equation}}
\newcommand{\ve}{\varepsilon}
\begin{document}
\comment[Comment on `The minimum-uncertainty squeezed states']{Comment on `The minimum-uncertainty squeezed states for atoms and photons in cavity'}
\author{A Jadczyk}
\address{Laboratoire de Physique Th\'{e}orique, Universit\'{e} de Toulouse III \& CNRS, 118 route de Narbonne, 31062 Toulouse, France}
\ead{ajadczyk@physics.org}
\begin{abstract}
In a recent work, Kryuchkov, Suslov and Vega-Guzm\'{a}n [20013 \jpb {\bf 46} 104007] described a multi-parameter family of minimum-uncertainty states satisfying the time-dependent Schr\"{o}dinger equation for the harmonic oscillator. We show how a different parametrization and a proper geometrical insight reduces a complicated set of equations to two simple circular motions.
\end{abstract}
The minimum-uncertainty squeezed states (non-classical states of light) are becoming increasingly important in physics, chemistry and technology. Their general theory is described, from several angles, in \cite{henry,dodonov,drummond}, while examples of application are given e.g. in \cite{clerk1,clerk2}. In a recent work, Kryuchkov, Suslov and Vega-Guzm\'{a}n \cite{suslov} described a multi-parameter family of minimum-uncertainty states satisfying the time-dependent Schr\"{o}dinger equation for the harmonic
oscillator in one dimension,
\begin{equation}
2i\psi _{t}+\psi _{xx}-x^{2}\psi =0.  \label{Schroudinger}
\end{equation}%
They defined the dynamics of the normalized wave packet by the following system of equations:
\begin{equation}
\psi \left( x,t\right) =e^{i\left( \alpha \left( t\right) x^{2}+\delta
\left( t\right) x+\kappa \left( t\right) +\gamma \left( t\right) \right) }%
\sqrt{\frac{\beta \left( t\right) }{\sqrt{\pi }}}\ e^{-\left( \beta \left(
t\right) x+\ve \left( t\right) \right) ^{2}/2},  \label{WaveFunction}
\end{equation}%
where%
\begin{eqnarray}
&&\alpha \left( t\right) =\frac{\alpha _{0}\cos 2t+\sin 2t\ \left( \beta
_{0}^{4}+4\alpha _{0}^{2}-1\right) /4}{\beta _{0}^{4}\sin ^{2}t+\left(
2\alpha _{0}\sin t+\cos t\right) ^{2}},  \label{hhA} \\
&&\beta \left( t\right) =\frac{\beta _{0}}{\sqrt{\beta _{0}^{4}\sin
^{2}t+\left( 2\alpha _{0}\sin t+\cos t\right) ^{2}}},  \label{hhB} \\
&&\gamma \left( t\right) =-\frac{1}{2}\arctan \frac{\beta
_{0}^{2}\tan t}{1+2\alpha _{0}\tan t},  \label{hhG} \\
&&\delta \left( t\right) =\frac{\delta _{0}\left( 2\alpha _{0}\sin t+\cos
t\right) +\ve _{0}\beta _{0}^{3}\sin t}{\beta _{0}^{4}\sin
^{2}t+\left( 2\alpha _{0}\sin t+\cos t\right) ^{2}},  \label{hhD} \\
&&\ve \left( t\right) =\frac{\ve _{0}\left( 2\alpha _{0}\sin
t+\cos t\right) -\beta _{0}\delta _{0}\sin t}{\sqrt{\beta _{0}^{4}\sin
^{2}t+\left( 2\alpha _{0}\sin t+\cos t\right) ^{2}}},  \label{hhE} \\
&&\kappa \left( t\right) =\sin ^{2}t\ \frac{\ve
_{0}\beta _{0}^{2}\left( \alpha _{0}\ve _{0}-\beta _{0}\delta
_{0}\right) -\alpha _{0}\delta _{0}^{2}}{\beta _{0}^{4}\sin ^{2}t+\left(
2\alpha _{0}\sin t+\cos t\right) ^{2}}  \label{hhK} \\
&&\qquad \quad +\frac{1}{4}\sin 2t\ \frac{\ve _{0}^{2}\beta
_{0}^{2}-\delta _{0}^{2}}{\beta _{0}^{4}\sin ^{2}t+\left( 2\alpha _{0}\sin
t+\cos t\right) ^{2}},\nonumber
\end{eqnarray}%
while the constants $\alpha _{0},$ $\beta _{0}\neq 0, $ $\delta _{0},$ $%
\ve _{0},$ are real-valued initial data with $\alpha_0, \beta_0\neq 0.$\\
This parametrization, while it does work, leads to unnecesarily complicated calculations, and obscures
an extremely simple dynamical mechanism that is behind the time evolution of squeezed states and their characteristic "breathing" behavior \cite{henry}. In the following we describe a much simpler dynamics
and parametrization of the same wavepacket.\\
Our primary parameters are: a point $q(t),p(t)$ in the phase space, and a complex number $\zeta(t),$ with  $|\zeta(t)|<1$ (Poincar\'{e} disk). Their simple dynamics, consisting of two circular motions,  is given by
\begin{eqnarray}
q(t)&=&q_0 \cos t +p_0 \sin t\\
p(t)&=&p_0\cos(t)-q_0\sin t,\label{qtpt}\\
\zeta(t)&=&e^{2it}\zeta_0,\quad |\zeta_0|<1\label{zt}.
\end{eqnarray}
Define (Cayley transform)
\be z(t)=u(t)+iv(t)=\frac{1+\zeta (t)}{1-\zeta(t)},\ee
\be u_0+iv_0=\frac{1+\zeta_0}{1-\zeta_0}. \label{cayley}\ee
Then $u_0$ is necessarily positive, $v_0$ is an arbitrary real initial data. These definitions entail
\begin{eqnarray}
u(t)&=&\frac{u_0}{u_0^2\sin^2 t+(\cos t -v_0\sin t)^2},\\
v(t)&=&\frac{2v_0\cos 2t +(1-u_0^2-v_0^2)\sin 2t}{2\left(u_0^2\sin^2 t+(\cos t-v_0\sin t)^2\right)}.
\end{eqnarray}
Define the following normalized time-dependent state
\be \psi_0(x,t)=\left(\frac{u(t)}{\pi}\right)^{\frac{1}{4}}\,e^{i\left(x-\frac{q(t)}{2}\right)p(t)-\frac{1}{2}z(t)\,\left(x-q(t)\right)^2}.\ee
Then, as it can be easily checked, the density matrix $\rho=|\psi\rangle\langle\psi|$ satisfies the time evolution equation with the harmonic oscillator Hamiltonian, and can be used for calculating probability distributions (including the Wigner distribution) and expectation values of all observables. For instance, the $x$-space probability distribution $P_c(x,t),$ the $p$-space probability distribution $P_p(p,t),$ and the Wigner phase space distribution $P(x,p,t)$ are easily seen to be given by
\be P_x(x,t)=\sqrt{\frac{u(t)}{\pi}}\,e^{-u(t)\left(q(t)-x\right)^2},\,P_p(p,t)=\sqrt{\frac{u(t)}{\pi}}\,\frac{e^{-\frac{u(t) (p-p(t))^2}{|z(t)|^2}}}{|z(t)|}.\ee
\be P(x,p,t)=\frac{e^{-\frac{\left(p-p(t)+v(t)(x-q(t)\right)^2+u(t)^2(q(t)-x)^2}{u(t)}}}{\pi}.\ee
It is to be noted that all three distributions become more symmetric in $x$ and $p$ when written in terms of polar coordinates $r,\theta$ of $\zeta.$
\be
P_x(x,t)=\frac{1}{\sqrt{\pi}}e^{-f_-^2(t)(x-q(t))^2},\quad P_p(p,t)=\frac{1}{\sqrt{\pi}}e^{-f_+^2(t)(p-p(t))^2},
\ee
\be P(x,p,t)=\frac{1}{\sqrt{\pi}}e^{-\frac{(x-q(t))^2}{f_-^2(t)}-\frac{(p-p(t))^2}{f_+^2(t)}+\frac{4r_0(x-q(t))(p-p(t)) \sin (\theta_0+2t)}{1-r_0^2}},\ee
where 
\be f_\pm(t)=\sqrt{\frac{1-r_0}{1+r_0^2\pm 2r_0 \cos (\theta_0+2t)}}.\ee
Sometimes, however, we may need to construct superpositions of different states, in which case we need not only a density matrix, but a state vector, where the overall phase is also important, Such a state vector satisfying the Schr\"{o}dinger equation is given by
\be \psi(x,t)=e^{i\phi(t)}\,\psi_0(x,t),\ee
where
\begin{eqnarray} \phi(t)&=&p_0q_0\cos^2 t+\frac{1}{4}(p_0^2-q_0^2)\sin 2t\nonumber\\
&-&\frac{1}{2}\arctan \left(\frac{u_0\tan t}{1-v_0\tan t}\right)+\frac{1}{2}v_0 q_0^2.
\end{eqnarray}
The function $\phi(t)$ above is determined uniquely (up to an additive constant) by the requirement that $\psi(x,t)$ satisfies the Schr\"{o}dinger equation.\\ 
We describe now the relation between the two parametrizations. From $(q,p,u,v)$ to $\alpha,\beta,\gamma,\varepsilon,\kappa$ (where, for simplicity, we skip the argument $t$ in all parameters)
\begin{eqnarray}
\alpha&=&-\frac{v}{2},\,\beta=\sqrt{u},\,\delta=p+vq,\\
\ve&=&-\sqrt{u}\,q,\,\kappa+\gamma=\phi-q\left(p+\frac{1}{2}qv\right).\end{eqnarray}
From $\alpha,\beta,\gamma,\varepsilon,\kappa$ to $(q,p,\zeta)$
\begin{eqnarray}
q&=&-\frac{\ve}{\beta},\,p=\delta-\frac{2\alpha\ve}{\beta},\,u=\beta^2,\\
v&=&-2\alpha,\,\phi=\kappa+\gamma+\frac{\ve(\alpha\ve-\beta\delta)}{\beta^2},\\
z&=&u+iv,\,\zeta=\frac{z-1}{z+1}.
\end{eqnarray}
We note that simultaneous change of signs of $\beta_0$ and $\ve_0$ does not change the wavepacket $\psi_0,$ therefore $\beta_0$ can be always chosen positive. Additionally, the condition $\alpha_0\neq 0$ imposed in \cite{suslov} is unnecessary.\\
The breathing phenomenon of the wavepacket, that is totally obscured in the parametrization of \cite{suslov} is clearly encoded in our parametrization in the rotation of the complex variable $\zeta$ with half of the period of the circular motion in the phase space. It is responsible for the rigid rotation around the origin of the Wigner distribution that produces
oscillating scaling when integrated over momenta. In polar coordinates, $\zeta=r\,e^{\theta},$ $r$ determines the eccentricity $e=4r/(1+r)^2$, while $\theta$ defines the angle of the squeezed state elliptical shape of the Wigner distribution (as shown in the attached {\it Mathematica} notebook \cite{mat}.  Finally we mention that the complex space of the variable $z=u+iv,\, u>0,$ responsible for ``breathing'' has been discussed by Erich K\"{a}hler in his studies of the geometry of quantization \cite[Raum-Zeit Individuum p. 661]{kahler} where he has called this space ``pneuma'' (he used the term ``soma'' for the phase space of $p$'s and $q$'s).

\end{document}